# Efficient Greenhouse Temperature Control with Data-Driven Robust Model Predictive

Wei-Han Chen and Fengqi You

*Abstract*— Appropriate greenhouse temperature should be maintained to ensure crop production while minimizing energy consumption. Even though weather forecasts could provide a certain amount of information to improve control performance, it is not perfect and forecast error may cause the temperature to deviate from the acceptable range. To inherent uncertainty in weather that affects control accuracy, this paper develops a data-driven robust model predictive control (DDRMPC) approach for greenhouse temperature control. The dynamic model is obtained from thermal resistance-capacitance modeling derived by the Building Resistance-Capacitance Modeling (BRCM) toolbox. Uncertainty sets of ambient temperature and solar radiation are captured by support vector clustering technique, and they are further tuned for better quality by training-calibration procedure. A case study that implements the carefully chosen uncertainty sets on robust model predictive control shows that the DDRMPC has better control performance compared to rule-based control, certainty equivalent MPC, and robust MPC.

## I. INTRODUCTION

### A. Greenhouse Climate

Maintaining greenhouse climate is an important factor to crop growth. Among several different environment conditions that should be considered such as temperature, light, carbon dioxide, and humidity, temperature and relative humidity are the two main factors that should be carefully controlled. Regulating temperature in an appropriate range can not only increase crop production but also prevent plants from heat stress or cold damage. Meanwhile, constraining relative humidity decreases the possibility of developing leaf mold, which may damage crops severely. Studies on greenhouse climate control have been performed by researchers, including nonlinear control methods, adaptive control, robust, optimal control, energy balance models, model predictive control [1-14].

### B. Model Predictive Control

Among various approaches, model predictive control (MPC) is an effective strategy that utilizes prediction of disturbances to optimize future system behavior under certain constraints [15]. At each time step, the controller solves an optimization problem based on a model that shows the relationship between system states, control inputs, and disturbances. Only the first control input is implemented while the rest is discarded. This process repeats for all time steps to derive control trajectories. MPC is an ideal framework for building control because building dynamics are slow and the system model incorporates disturbances and constraints can be derived from first principles models [16]. Another advantage of MPC on greenhouse climate control is that a greenhouse can usually be considered as a large room. In this way, the model is easier to be obtained in contrast to a building with multiple rooms which involve number of system states is large. The benefits of MPC on building control have been demonstrated in comparison to conventional methods such as rule-based control (RBC). Numerous studies on greenhouse climate control that adopt the MPC have been investigated [4, 5, 7, 13, 17].

Despite the various advantages of MPC on greenhouse control, it is impossible to perfectly predict weather, which is the major disturbance in greenhouse control problems. For example, weather prediction of ambient temperature may deviate from the true measurement. The uncertainty of disturbances might cause system states to violate specified constraints and damage crop production. In order to cope with uncertainties, robust MPC (RMPC) can be adopted. When uncertainty is bounded, RMPC could ensure that system states would not violate the constraints even when the worst-case scenario occurs. The control inputs may end up more expensive to compensate for robustness. RMPC are implemented on greenhouse in some studies [17].

RMPC may lead to over-conservative results, which is not favorable. Although it is guaranteed that constraints in RMPC would not be violated in the worst-case scenario, the probability of such scenario to happen could be excessively low. To prevent extreme cases, more expensive control inputs are required, which would lead to a waste. In this work, a data-driven RMPC (DDRMPC) framework for greenhouse temperature model is proposed to reduce the conservatism. Firstly, the state-space model of the greenhouse is generated based on building elements construction. Secondly, historical weather forecast data and historical weather measurement data are gathered. These two sets of data could generate uncertain forecast errors. High-density region of the uncertain prediction errors is captured by a machine learning technique, support vector clustering (SVC). The historical data information can then be incorporated into RMPC, and conservatism is reduced. To solve the optimization problem in DDRMPC, affine disturbance feedback (ADF) policy is utilized to provide tractable approximations. Lastly, the optimization problem can be solved effortlessly by off-the-shelf solvers when the robust counterpart has been derived.

The contributions of this paper are summarized below:

- A DDRMPC framework for greenhouse temperature control using SVC is constructed;
- A simulation of greenhouse temperature control based on real weather data demonstrates better control

W.-H. Chen and F. You are with the Robert Frederick Smith School of Chemical and Biomolecular Engineering, Cornell University, Ithaca, NY 14853 USA (e-mail: wc593@cornell.edu; fengqi.you@cornell.edu).

performance of DDRMPC comparing to other conventional methods.

*C. Outline*

The structure of the paper is as follows: Section II presents the construction of greenhouse temperature model. In Section III, different control strategies are set up for controlling greenhouse temperature. In Section IV, a simulation of controlling temperature in a greenhouse located at New York City, USA is served as a case study of the proposed model and control strategies, and the simulation results of different control strategies are discussed. Finally, Section V summarizes major conclusions and future works.

## II. MODEL FORMULATION

*A. Greenhouse Model Formulation*

In greenhouse MPC, a model is required for predicting greenhouse climate (e.g. temperature, relative humidity, $CO_2$ concentration) as a function of control inputs (e.g. HVAC system heating power) and disturbances (e.g. ambient temperature) so that the climate can be constrained in a satisfied range. The cost function and constraints of control inputs are also required to be modeled.

Models can be categorized in two groups: first-principles models and black box models. The former ones are derived according to physical equations which require building geometry (e.g. room sizes, wall thickness), building construction (e.g. wall materials), and systems data (e.g. location). The latter ones are obtained by capturing the relationships between input (control inputs and disturbances) and output (climate states) data. In comparison to the latter ones, the former ones are more comprehensible and could be easily constructed for different buildings only by varying parameters.

The Building Resistance-Capacitance Modeling (BRCM) MATLAB toolbox [18] uses first-principles models to derive building models which are specially designed for MPC. Although there are several other programs that provide building simulation (e.g. EnergyPlus, TRNSYS), most of them provide highly nonlinear models and their complexities are more than what is needed for MPC. In fact, BRCM toolbox shows that its simplified building model results in acceptable prediction difference from measurement or EnergyPlus model, which is about 0.5-1 °C after several days of prediction. Therefore, utilizing BRCM toolbox to develop building model could significantly reduce the modeling effort without losing significant accuracy.

The main approach of BRCM toolbox is thermal resistance-capacitance modeling. Building elements are served as resistances and capacitances so the model is an analogy to electrical circuit modeling where temperature corresponds to voltage, heat flux to current, thermal capacitance to electrical capacitance, and thermal resistance to electrical resistance. Thermal conduction, convection, and radiation are considered under the law of energy conservation. Hence, the model is described by linear ordinary differential equations. Further after discretized, the state space representation is given by

$$x_{k+1} = Ax_k + B_u u_k + B_v v_k \tag{1}$$

where $x_k \in \mathbb{R}^n$ is the state, $u_k \in \mathbb{R}^m$ is the control input, and $v_k \in \mathbb{R}^p$ is the weather disturbance at time step $k$, respectively. The matrices $A$, $B_u$, and $B_v$ are of appropriate sizes.

As prediction error is unavoidable in weather forecast, $w_k$ is implemented to represent prediction error. The model then becomes

$$x_{k+1} = Ax_k + B_u u_k + B_v v_k + B_w w_k \tag{2}$$

where system matrix $A$, $B_u$, $B_v$, and $B_w$ is of appropriate size.

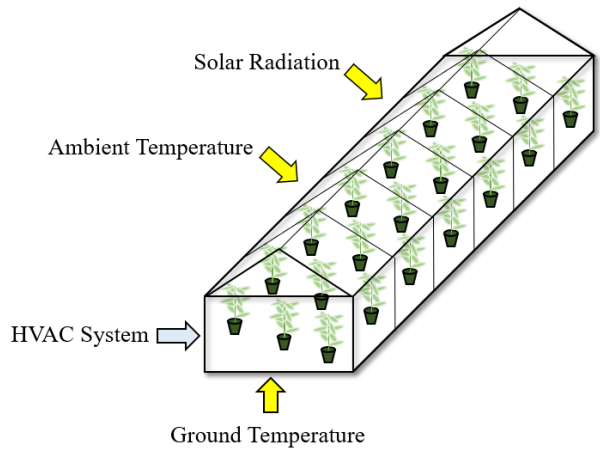

Figure 1. Greenhouse structure model.

In this work, the states we consider are greenhouse temperature, floor temperature, ceiling temperature, and wall temperature. The control input is the heating power of HVAC system. The disturbances are solar radiation, ambient temperature, and ground temperature. We assume ground temperature is perfectly known so prediction errors are only for solar radiation and ambient temperature [19]. The structure of greenhouse model including input and disturbances is shown in Fig. 1.

*B. Uncertainty Set Formulation*

An uncertain disturbance $w$ can be bounded by an uncertainty set $W$, which is an important factor for RMPC. If $W$ is chosen too small, protection would be inadequate and may lead to constraint violations. On the other hand, if $W$ is chosen too large, the control trajectory would be over-conservative, and more inputs may be added to protect the extreme case. Therefore, the uncertainty set $W$ should be chosen precisely.

Before constructing an uncertainty set for ambient temperature prediction error $w_{temp}$, pairs of historical forecast data and historical measurement are required. Samples of temperature prediction errors can then be calculated from $w_{temp} = \tilde{v}_{temp} - \hat{v}_{temp}$ where $\tilde{v}_{temp}$ is the historical measurement for ambient temperature, and $\hat{v}_{temp}$ is the historical forecast. To obtain the uncertainty set from these $N$ samples, we adopt SVC [20], which tries to find the radius of the minimal sphere that can capture data without considering outliers. Weighted generalized intersection kernel (WGIK) proposed in [21] is implemented when solving the dual form of SVC optimization problem. Unlike some common kernels (e.g. RBF, polynomial) which would cause a burden when solving robust optimization, WGIK is especially suited for robust optimization due to its linearity. The data-driven uncertainty set is shown as

$$w_{temp} \in D_{temp} = \left\{ w_{temp} \left| \sum_{i \in SV} \alpha_i \left\| \mathbf{Q}(w_{temp} - w_{temp}^{(i)}) \right\|_1 \leq \theta \right. \right\} \quad (3)$$

where $\mathbf{Q}$ is a weighting matrix that can be obtained from the covariance matrix of $w_{temp}$. Model parameters $\{\alpha_i\}$ and uncertainty set parameters $\theta$ are determined after solving the dual form of SVC using WGIK. Since (3) is a polytope, solving the robust optimization problem with this uncertainty set could be accomplished without difficulties.

It is basically the same procedure for constructing solar radiation uncertainty set as temperature uncertainty set. The only difference is that historical solar radiation forecast is hard to obtain. In order to tackle this problem, we adopt the following simple equations which estimate solar radiation from cloud coverage since historical cloud coverage data is fairly easy to obtain comparing to historical solar radiation forecast data [22].

$$v_{sol} = R_0(1 - 0.75(n/8)^{3.4}), \quad (4)$$

$$R_0 = 990 \sin \frac{\phi_{tp} - \phi_p}{2} - 30, \quad (5)$$

where $R_0$ is clear sky insolation, $n$ is the cloud coverage ranging from 0 to 1, $\phi_{tp}$ is the hour solar elevation angle at the beginning of the time step, and $\phi_p$ is the hour solar elevation angle at the end of the time step.

Since solar radiation forecast data are obtained, solar radiation prediction errors are calculated from $w_{sol} = \tilde{v}_{sol} - \hat{v}_{sol}$ where $\tilde{v}_{sol}$ is the historical measurement for solar radiation, and $\hat{v}_{sol}$ is the historical forecast for solar radiation. The SVC-based solar radiation uncertainty set is shown as

$$w_{sol} \in D_{sol} = \left\{ w_{sol} \left| \sum_{i \in SV} \alpha_i \left\| \mathbf{Q}(w_{sol} - w_{sol}^{(i)}) \right\|_1 \leq \theta \right. \right\}, \quad (6)$$

which is in the same form as the SVC-based ambient temperature uncertainty set.

*C. Performance Guarantee and Uncertainty Set Tuning*

Since uncertainty sets are obtained from data-driven approach, we would like to ensure that our data-driven uncertainty set has appropriate probabilistic guarantee, or else the performance could not be guaranteed. The constructed data-driven uncertainty set $D$ is called $(1-\epsilon)$-prediction set if $\mathbb{P}\{\mathbf{w} \in D\} \geq 1-\epsilon$. Besides that, large confidence of uncertainty set $D$ being a $(1-\epsilon)$-prediction set is further demanded because constructing uncertainty set $D$ itself is random. This can be expressed as

$$\mathbb{P}_D \{\mathbb{P}\{\mathbf{w} \in D\} \geq 1-\epsilon\} \geq 1-\beta \quad (7)$$

where $\epsilon$ and $\beta$ are parameters that can be specified by our own.

Training-calibration procedure proposed by [23] is implemented to reach performance guarantee (7). All data are split into training data set and calibration data set. The uncertainty set can be written as

$$D = \{\mathbf{w} | f(\mathbf{w}) \leq \theta\} \quad (8)$$

where $f(\cdot)$ is the left-hand side of the constraint in (3) and (6), and the value of $\theta$ is tuned according to the calibration data set $\{\mathbf{w}_{calib}^{(i)}\}_{i=1}^{N_{calib}}$ following

$$\theta = \max_{1 \leq i \leq N_{calib}} \left\{ f\left(\mathbf{w}_{calib}^{(i)}\right) \right\}. \quad (9)$$

Also, if the number of calibration data follows

$$N_{calib} \geq \frac{\log \beta}{\log(1-\epsilon)}, \quad (10)$$

the calibrated SVC-based uncertainty set admits performance guarantee (7) [24].

### III. CONTROL STRATEGIES

In this work, different control strategies are simulated to compare their control performances. These are RBC, certainty equivalence MPC (CEMPC), RMPC, and the proposed DDRMPC.

*A. RBC*

RBC is the standard approach to control building temperature. The control inputs are triggered once the specified conditions are met. Although there are several complicated conditions that can be set up, they require a large amount of effort to tune RBC to perform well. The simple if-else condition we consider here is to turn on heating system at constant heating power C once the difference between the greenhouse temperature and the target $T$ of that time falls below a threshold $\delta$. During daytime, the greenhouse favors higher temperature than during nighttime. Therefore, the target of daytime is higher.

$$u_t = \begin{cases} C, & \text{if } x_t \leq T_t + \delta \\ 0, & \text{if } x_t > T_t + \delta \end{cases} \quad (11)$$

*B. CEMPC*

CEMPC takes predicted disturbance as true disturbance. When the prediction horizon $H$ is given, (1) can be written in a compact form

$$\mathbf{x} = \mathbf{A} x_0 + \mathbf{B}_u \mathbf{u} + \mathbf{B}_v \mathbf{v}. \quad (12)$$

The constraints are defined for control inputs and system states throughout the entire prediction horizon $H$. We assume that the control input $u$ should be between zero and a maximum value limited by…, which is $0 \leq u_t \leq u_{max}$. The greenhouse temperature should be above the target, which varies according to daytime or nighttime as $x_t \geq T_t$. The compact form is shown as

$$\mathbf{F}_x \mathbf{x} \leq \mathbf{f}_x, \quad \mathbf{F}_u \mathbf{u} \leq \mathbf{f}_u \tag{13}$$

where $\mathbf{F}_x$, $\mathbf{f}_x$, $\mathbf{F}_u$, and $\mathbf{f}_u$ are of appropriate sizes.

Different cost functions can be adopted, such as quadratic cost, linear cost, and probabilistic cost. A linear cost function is implemented in this work as

$$J_H = \sum_{t=0}^{H-1} c^T \cdot u_t \tag{14}$$

where $c$ is the cost of different control variables. However, the heating system is the only control variable in this work. Therefore, we could directly observe the energy consumption without any scaling required. Thus, $J_H$ can be viewed as a total amount of energy consumption by the heating system.

### C. RMPC

RMPC guarantees constraint satisfaction for the worst case of the bounded disturbances [25]. In order to ensure the tractability of the RMPC problem, ADF policy is adopted, and control input $u_t$ is parameterized according to the past disturbances as follows [26]

$$u_t = h_t + \sum_{j=0}^{t-1} M_{t,j} w_j. \tag{15}$$

It also can be written in a compact form as

$$\mathbf{u} = \mathbf{h} + \mathbf{M}\mathbf{w}$$

where

$$\mathbf{M} = \begin{bmatrix} 0 & \cdots & \cdots & 0 \\ M_{1,0} & 0 & \cdots & 0 \\ \vdots & \vdots & \ddots & \vdots \\ M_{H,0} & M_{H,1} & \cdots & 0 \end{bmatrix}, \quad \mathbf{h} = \begin{bmatrix} h_0 \\ h_1 \\ \vdots \\ h_H \end{bmatrix} \tag{16}$$

becomes decision variables that should be solved to determine the control inputs.

We assume the uncertainty set for RMPC to be $L_1$-norm-based as follows [27]

$$\mathbf{w} \in W = \{\mathbf{w} \mid \|\mathbf{w}\|_1 \leq \Omega\} \tag{17}$$

where $\Omega$ is the budget parameter that can adjust the conservatism. When $\Omega$ is larger, the uncertainty set becomes bigger. The RMPC problem can now be solved easily after the ADF policy is adopted.

### D. DDRMPC

DDRMPC adopts SVC to construct uncertainty sets that could tackle outliers. Furthermore, the performance guarantee is ensured after tuning uncertainty sets by the calibration data set. The approach to solving the optimization problem in DDRMPC also uses ADF policy shown as

$$\min_{\mathbf{M},\mathbf{h}} \mathbf{c}^T \mathbf{h}$$

$$\text{s.t. } \mathbf{F}_x \left[ \mathbf{A} x_0 + \mathbf{B}_u \mathbf{h} + \mathbf{B}_v \mathbf{v} + (\mathbf{B}_u \mathbf{M} + \mathbf{B}_w) \mathbf{w} \right] \leq \mathbf{f}_x, \forall \mathbf{w} \in D \tag{18}$$

$$\mathbf{F}_u \left[ \mathbf{M}\mathbf{w} + \mathbf{h} \right] \leq \mathbf{f}_u, \forall \mathbf{w} \in D$$

where $x_0$ is the initial state, and (18) is a convex optimization problem that can be solved after using robust optimization techniques.

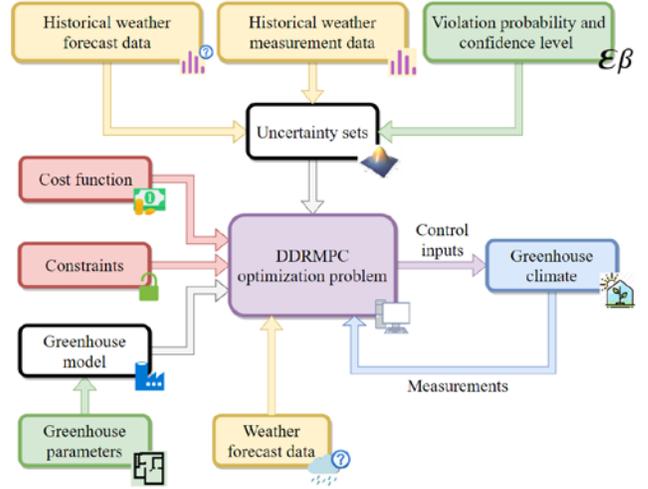

Figure 2. Schematic of DDRMPC implementation in greenhouse climate control.

## IV. CASE STUDY

### A. Problem Description

In this work, a greenhouse located at Brooklyn, NY, USA is simulated for closed-loop temperature control under different control strategies. The dimension of the greenhouse is 40 m × 13 m × 4 m. The material for roof and walls is 10 mm twin-wall polycarbonate which provides good insulation against heat. The floor is made of concrete. The disturbances considered are solar radiation, ambient temperature, and ground temperature. To control the room temperature, HVAC system is implemented. Historical weather forecast data and historical weather measurement data from January 2018 to June 2018 are collected from Meteogram Generator [28]. Historical solar radiation forecast data are hard to access. Therefore, the simple prediction model shown in (7) is adopted to estimate solar radiation from cloud coverage.

The temperature control is for tomatoes to grow in the greenhouse. Therefore, the control goal for daytime and night time is treated differently. From 6 am to 10 pm, the greenhouse temperature should be above 25 °C, and in the rest of time, the greenhouse temperature should be above 18 °C. Ground temperature is assumed to be a constant (18 °C). The maximum heating power for HVAC system is set as 300,000 W. For RBC, constant heating power $C$ is set as 60,000 W, and the threshold $\delta$ is 7 °C. For CEMPC, RMPC, and DDRMPC, the prediction horizon $H$ is 5 intervals, and the sampling interval is 1 hour. For RMPC, budget parameter $\Omega$ ranges from 0 to 6 to reveal different levels of conservatism. For DDRMPC, another set of historical weather data from January 2017 to June 2017 is obtained for constructing data-driven uncertainty set. The maximal violation probability and confidence level are set as $\epsilon = 0.05$ and $\beta = 0.10$, leading to $N_{\text{calib}} = 45$ samples. The schematic of a simulating implementation of DDRMPC is shown in Fig. 2. The performance bound, which solves deterministic MPC with perfect weather forecast, is implemented as a benchmark.

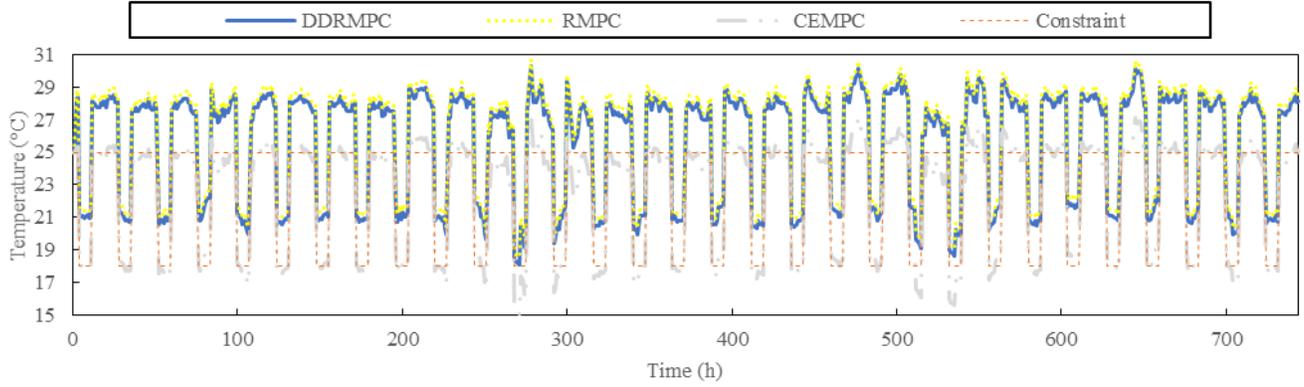

Figure 3.  Greenhouse temperature profile in January 2018.

All of the optimization problems are computed by YALMIP with the solver GUROBI [29, 30]. The greenhouse temperature control is simulated from January 2018 to June 2018 on a computer with Intel Core i7-6700 processor at 3.40 GHz and 32 GB of RAM.

*B. Results and Discussions*

The results of different control strategies under each month are simulated, and Fig. 3 is an example that presents the temperature profile of January. The profile has a diurnal cycle, with higher temperature during daytime. In all cases, RMPC and DDRMPC leave some margins from the temperature constraints while CEMPC violates the constraints frequently, which demonstrates the robustness of RMPC and DDRMPC.

CEMPC shows the least conservative control profile. Although CEMPC consumes the least heating power shown in Table 2, it ends up with the most constraints violations in Table 1. Assuming prediction as true value makes CEMPC violate constraints. Whenever the weather is colder than predicted, CEMPC would violate the constraint. Some violations are so severe that the temperature drops to 15 °C. This is undesirable in a greenhouse because crops are usually sensitive to temperature. Even a 2-3 °C difference may cause serious damages to crop.

RMPC is tuned according to the budget parameter $\Omega$. When $\Omega = 0$, RMPC is the same as CEMPC which does not consider uncertainty. When $\Omega = 6$, RMPC results in no more constraint violation. To achieve this result, RMPC requires more heating power than CEMPC and DDRMPC. However, DDRMPC also results in nearly zero constraint violation.

Table 3 shows the trade-off between violation and energy consumption. Energy consumption comparison of different control strategies are on the same basis of performance bound (PB) result. It reveals that although CEMPC induces the least energy consumption, it violates constraints frequently. RBC results in more energy consumption and constraint violation percentage than RMPC and DDRMPC, which shows that it performs worse than those two approaches. While DDRMPC violates constraints slightly, it consumes less energy than RMPC. The violation percentage only tells how frequent a control strategy violates constraints but does not show how serious the violation is. In the third row, another indicator is added to compare how serious the violation is for different control strategies. Violation amount calculates the area below the violation, and it has similar results to violation percentage, which CEMPC violates a great amount, and RBC performs worse in this case as well. Although DDRMPC violates 24 Kh in total, it does not affect much when the total simulation time of 3,624 h is considered.

The benefit of DDRMPC is revealed since it can achieve similar constraint violation while consuming less energy consumption. By capturing the distribution of weather uncertainty, DDRMPC could get rid of outliers that only happen in extreme cases. In addition, unlike RMPC, which uncertainty sets are in a fixed shape, DDRMPC implements uncertainty sets that change shapes according to high-density region of data, which could further reduce conservatism.

RBC requires fine tuning to achieve better performance. In our simple setup, even when RBC consumes similar energy as DDRMPC, it still violates constraints quite often. This shows another advantage of DDRMPC, which is the convenience of tuning the conservatism. Adjusting parameters of violation probability and confidence level is enough for DDRMPC.

TABLE I. CONSTRAINT VIOLATION PERCENTAGE IN EACH MONTH

|  | Jan. | Feb. | Mar. | Apr. | May |
|---|---|---|---|---|---|
| RBC (%) | 23.39 | 6.99 | 6.45 | 4.58 | 3.61 |
| CEMPC (%) | 57.80 | 46.73 | 44.35 | 54.31 | 43.47 |
| RMPC (%) | 0 | 0 | 0 | 0 | 0 |
| DDRMPC (%) | 0 | 0 | 0 | 0.42 | 0 |

TABLE II. HEATING POWER USAGE IN EACH MONTH

|  | Jan. | Feb. | Mar. | Apr. | May |
|---|---|---|---|---|---|
| PB (MWh) | 34.24 | 24.50 | 27.87 | 21.10 | 10.02 |
| RBC (MWh) | 40.98 | 34.19 | 38.86 | 33.30 | 24.35 |
| CEMPC (MWh) | 33.98 | 24.64 | 28.10 | 20.89 | 10.06 |
| RMPC (MWh) | 41.94 | 31.83 | 36.05 | 28.57 | 17.30 |
| DDRMPC (MWh) | 40.93 | 30.90 | 35.03 | 27.59 | 16.34 |

TABLE III. TRADE-OFF BETWEEN ENERGY USAGE AND VIOLATION INDICATORS

|  | RBC | CEMPC | RMPC | DDRMPC |
|---|---|---|---|---|
| Energy usage* (%) | 46 | 0 | 32 | 28 |
| Violation percentage (%) | 9.0 | 49 | 0 | 0.084 |
| Violation amount (Kh) | 560 | 1,923 | 0 | 24 |

* Additional energy usage in % of PB

## V. CONCLUSION

A data-driven robust model predictive control (DDRMPC) was applied to greenhouse temperature control in this paper. Uncertainty sets for temperature and solar radiation were constructed by adopting the SVC approach on historical data. The simulation of a greenhouse in Brooklyn showed that DDRMPC had better control performance compared to RBC, CEMPC, and RMPC. For future work, more complicated climate models can be considered. For example, humidity is an important state in greenhouse climate that also has strong relation with temperature. In this way, a more comprehensive greenhouse model can be formed.